\documentclass[prb, twocolumn, superscriptaddress]{revtex4-2}
\usepackage[margin= 0.55in,top= 15.0mm]{geometry}
\usepackage{amsmath,amssymb,bm}
\usepackage{hyperref}
\usepackage{graphicx}
\usepackage{epstopdf}
\usepackage{latexsym}
\usepackage[usenames, dvipsnames]{color}
\usepackage[usenames, dvipsnames]{xcolor}
\usepackage{braket}
\usepackage{float}
\usepackage[normalem]{ulem}
\usepackage{comment}
\usepackage{mathtools}
\usepackage{array}
\usepackage{tabu}
\usepackage{multirow}
\usepackage[inline]{enumitem}
\usepackage{cleveref}
\usepackage{xcolor}
\usepackage[normalem]{ulem}
\usepackage{natbib}
\usepackage{amsmath}

\begin{document}
\title{Cooperative motion in equilibrium phases across two-dimension melting in pure and disordered systems}
\author{Saikat Dutta}
\affiliation{Department of Physical Sciences, Indian Institute of Science Education and Research (IISER) Kolkata, Mohanpur - 741246, West Bengal, India}
\author{Prashanti Jami}
\affiliation{Department of Physical Sciences, Indian Institute of Science Education and Research (IISER) Kolkata, Mohanpur - 741246, West Bengal, India}
\author{Pinaki Chaudhuri}
\affiliation{The Institute of Mathematical Sciences, Taramani, Chennai 600113, India}
\author {Chandan Dasgupta}
\affiliation{Department of Physics, Indian Institute of Science, Bangalore 560012, India}
\affiliation{International Centre for Theoretical Sciences, Tata Institute of Fundamental Research, Bangalore 560089, India}
\author{Amit Ghosal}
\affiliation{Department of Physical Sciences, Indian Institute of Science Education and Research (IISER) Kolkata, Mohanpur - 741246, West Bengal, India}

\begin{abstract}
We uncover the dynamics of particles with Gaussian core interactions across melting in pure and disordered two-dimensional (2D) systems. Intriguing signatures of cooperative motion of particles in string-like paths are found at low temperatures. Such a motion, while common to glasses and supercooled liquids, are realized here in traditional equilibrium phases, including in pure systems. We explore the interplay of such motion and impurities and report their repercussions on spatiotemporal correlations. In particular, cooperative motion seems to cause a departure from the diffusive dynamics, causing slow relaxation.
\end{abstract}

\maketitle

\textit{Introduction:}
The dynamics in a many-particle system characterize its phases at different temperatures ($T$) because they respond to inter-particle interactions and single-particle potentials.
Particles undergo collective oscillations (phonons) in a solid, resulting in a Gaussian distribution of their displacements~\cite{ashcroft1976solid}. In liquids, conversely, particles move diffusively, also giving rise to Gaussian displacements, the width of which, however, is significantly broader~\cite{landau2013fluid, kundu2015fluid}. A deviation from the Gaussian form signals complex motion and is rife in glassy phases~\cite{Cipelletti_2005,RevModPhys.83.587,ediger2000spatially}.

Melting in two-dimension (2D) is special -- Mermin-Wagner theorem~\cite{PhysRevLett.17.1133, PhysRev.176.250} prohibits a solid in 2D at finite $T$ with true long-range positional order.
A two-step melting in a pure 2D system still occurs as described by the Berezinskii-Kosterlitz-Thouless-Halperin-Nelson-Young (BKTHNY) theory~\cite{berezinskii1972destruction, Kosterlitz:1973xp, PhysRevB.19.2457, PhysRevLett.41.121,PhysRevB.19.1855}, leaving a hexatic phase between the solid and liquid. Within BKTHNY theory, 2D meltings transpire by the unbinding of thermal defects. The hexatic phase is promoted by unbinding dislocation pairs disrupting the quasi-long range positional order, which depletes the shear modulus~\cite{RevModPhys.60.161,chaikin1995principles}.
In the next step, isolated dislocations decompose into free disclinations, degrading the bond-orientation~\cite{PhysRevB.19.2457} yielding an isotropic liquid.
In a recent study~\cite{PhysRevE.109.L062101}, we uncovered how impurities inherent to any real system modify this scenario 
by turning this into an inhomogeneous single-step melting.
Addressing the dynamical responses we report in this letter, the fascinating signatures of dynamical heterogeneity at low $T$ in clean and disordered systems arise from the spatially correlated motion of particles along long, tortuous paths.
The resulting distribution of displacements, even in pure systems, becomes non-Gaussian and causes slow relaxation akin to glassy systems.

 The effects of impurities on 2D melting have been addressed in recent experiments~\cite{PhysRevLett.111.098301,PhysRevB.93.144503,singh2022observation}. The anisotropic dynamics of vortices of a Type-II superconductor along string-like `motion paths' have been reported~\cite{PhysRevB.108.L180503}. The dynamics of melting were investigated for colloidal particles~\cite{PhysRevLett.82.2721, PhysRevLett.85.3656}. The cooperative motion of particles is a common characteristic of glassy dynamics~\cite{PhysRevLett.80.2338, nagamanasa2011confined}. They are also found in recent simulations~\cite{maccari2024fragiletostrongglasstransitiontwodimensional, zangi2004cooperative, kim2013simulation, Shiba_2009}. We analyze below the cooperative dynamics and their consequences in the presence of impurities.

{\textit{Model \& method:} 
We use the Gaussian-core model (GCM) \cite{gcm,PhysRevLett.106.235701} of repulsion between classical particles. This pairwise interaction between particles at a distance $r$ is given by $V(r) = \exp\left(-r^{2}/\sigma^{2}\right)$. Here $\sigma$ signifies the `screening length' of the interaction. Beyond its relevance towards
phenomenology~\cite{likos2001effective},
a pure system of GCM particles demonstrates a two-step BKTHNY melting~\cite{PhysRevLett.106.235701, PhysRevE.109.L062101}.
We consider a 2D system with periodic boundary conditions with $N=4096$ particles. Two distinct types of disorder are investigated: Random pinning (RP) and commensurate pinning (CP). RP involves freezing a given fraction, $n_{\rm imp}$, of particles randomly from a high-temperature equilibrium liquid-like configuration. In contrast, CP freezes $n_{\rm imp}$ fraction of particles at randomly selected vertices of a perfect triangular lattice -- the ground state configuration of the pure system. These immobile particles serve as impurities. While CP introduces correlated disorder with long-range positional correlations, RP causes nearly uncorrelated disorder, though short-range liquid-like correlations exist~\cite{PhysRevE.109.L062101, PhysRevLett.111.098301, PhysRevE.92.032110}. These systems are investigated with $N=4356$ particles, with $n_{\rm imp}= 3.5\%$, maintaining the particle density the same as in the pure and pinned systems for fair comparisons. We use Molecular dynamics simulation similar to Ref.~\onlinecite{PhysRevE.109.L062101} (See section I in supplementary materials (SM) ~\cite{SupplementalMaterial}). Results from pinned systems were averaged over $10$ independent pinning realizations.
Our tuning parameter, $T$, is expressed in terms of a dimensionless coupling $\Gamma^{-1} \sim T$ (see section I in SM ~\cite{SupplementalMaterial}).
We describe our key results below, beginning with the time-dependent Lindemann parameter.
\begin{figure*}[t]
\includegraphics[width= 1\textwidth]{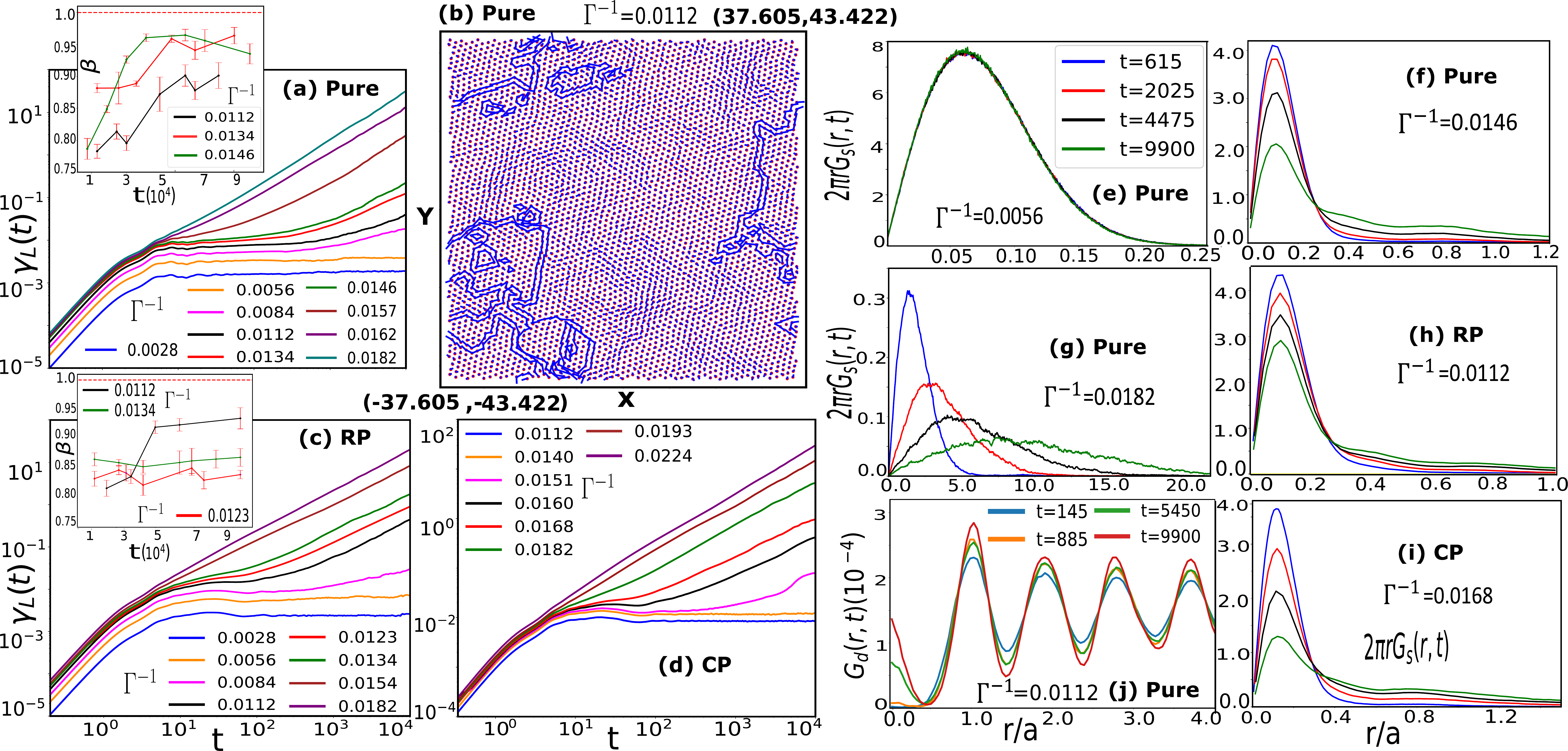}
\caption{{\bf The time-evolution of Lindemann parameter, van Hove correlations and underlying cooperative dynamics:} $\gamma_{L}(t)$ at different temperatures ($\Gamma^{-1}$) are shown for pure (a), RP (c) and CP (d) systems.
In addition to solid (constant $\gamma_{L}(t)$) and liquid phases ($\gamma_{L}(t) \sim t$) the pure and CP-systems feature an intermediate $T$-window (termed CMTR) where $\gamma_{L}(t) \sim t^{\beta(T)}$ with $\beta(T)<1$ at long times. For RP-systems, CMTR engulfs the entire low-$T$ phase until melting to the liquid. The insets of panels (a) and (c) display $\beta(T)$ of versus $t$ within CMTR for pure and RP-systems. 
Panel (b) describes displacements of particles in the pure system at $\Gamma^{-1}=0.0112$ until $t\le 10^{4}$. The red dots indicate the initial positions of the particles. In contrast, the blue lines join the initial and final positions of individual particles and thereby represent their displacement, and the coordinates represent the size of the simulation box.
Panel (e) displays the $r$-dependence of $2\pi r G_{s}(r,t)$ for the pure system at a low $T$ for four distinct times, demonstrating the solid-like behavior. A similar trend was found for CP-systems, see Fig.S3(b) in SM~\cite{SupplementalMaterial}.
Panel (f) depicts the behavior of $2 \pi r G_{s}(r,t)$ in a pure system in CMTR ($\Gamma^{-1}=0.0112$). Notably, the peak of $G_{s}(r,t)$ does not alter its position; its height reduces progressively by producing long tails. Panel (g) shows the behavior of $2 \pi r G_{s}(r,t)$ in a pure system at high $T$ ($\Gamma^{-1}=0.0182$), displaying liquid-like behavior featuring progressive long tails by reducing the peak height, as well as by shifting its position. Such trends at large $T$ are nearly identical for Pure, RP, and CP-systems. 
Panels (h, i) show $2 \pi r G_{s}(r,t)$ in CMTR for RP- and CP-systems, respectively. The CMTR extends down to the lowest $T$ in RP-systems, and corresponding $G_{s}(r,t)$ is shown in Fig. S3(a) of SM~\cite{SupplementalMaterial}. Panel (j) depicts the $r$-dependence of $G_{d}(r,t)$ in CMTR of the pure system. The rising peak at $r=0$ with $t$, absent at $t=0$, signals the particles' cooperative motion.
}

\label{f1}
\end{figure*}

\begin{figure*}[t]
\includegraphics[width= 1\textwidth]{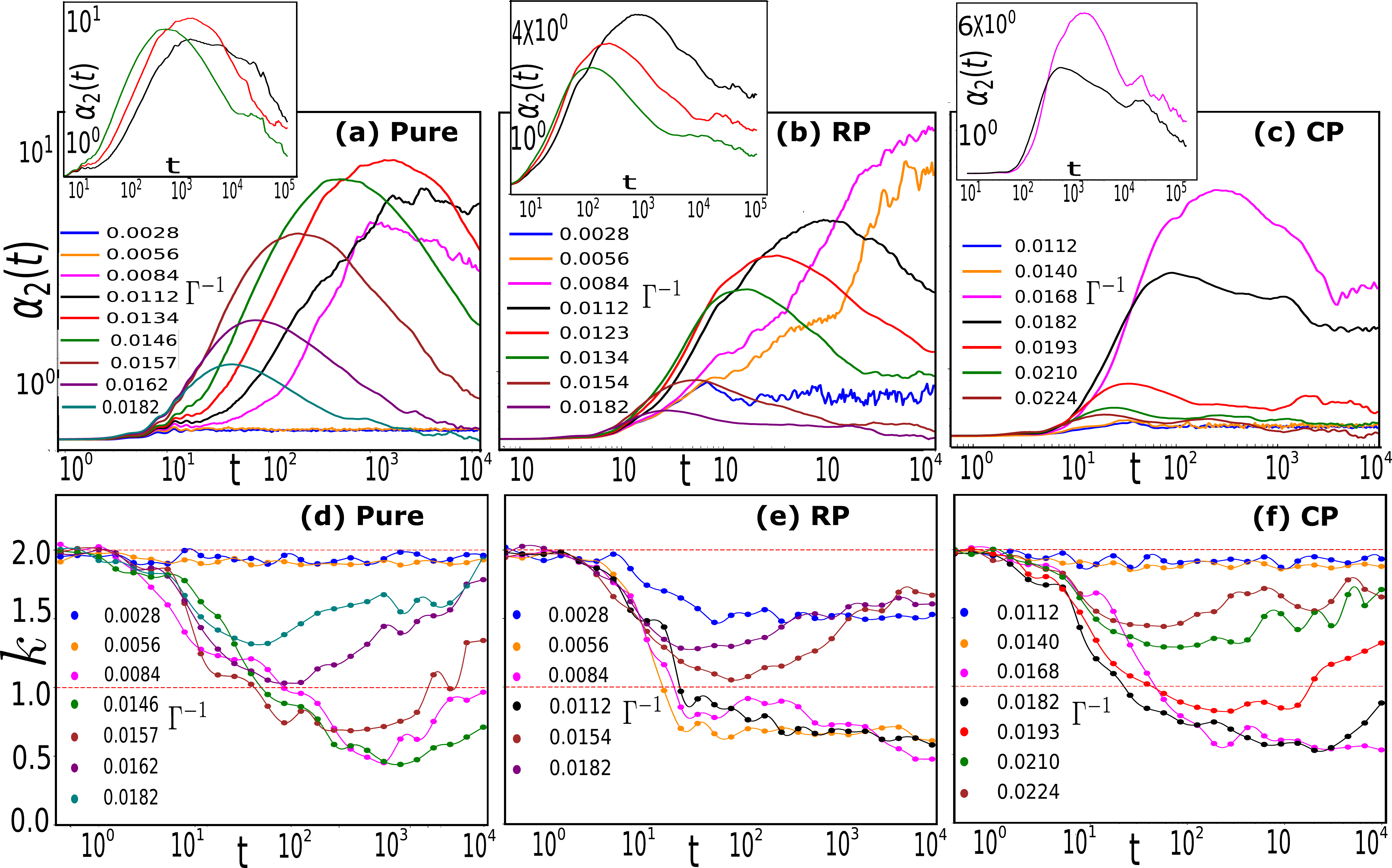}
\caption {{\bf The variation of the Non-Gaussian parameter} is shown in panels (a,b,c) for different $T \sim \Gamma^{-1}$ for pure, RP, and CP systems, respectively. The insets of panels (a),(b) and (c) illustrate the long-time ($t \leq 10^{5}$) behavior of $\alpha_2(t)$ in CMTR for pure-, RP- and CP- systems. The degree of non-Gaussian evolution of $G_{s}(r,t)$ is analyzed in panels (d), (e), and (f) for pure, RP, and CP systems, respectively, by studying the exponent $k(t)$ (see text)}. 
\label{f3}
\end{figure*}

\textit{Dynamic Lindemann parameter, $\gamma_{L}(t)$} is defined as the mean-square displacement of particles normalized appropriately by the lattice spacing $a$ of the underlying triangular lattice at $T=0$, i.e., $\gamma_{L}(t)=\langle (\Delta r(t))^{2}\rangle/2a^{2}$. Here, $\Delta r(t)$ is the cage-relative displacements (CRD), i.e., displacements of the particles relative to their first neighbors~\cite{BEDANOV1985289,peierls1979surprises,PhysRevLett.85.3656}(See section II in SM~\cite{SupplementalMaterial}). The significance of CRD, in contrast to standard displacement, is emphasized in Ref.~\onlinecite{illing2017mermin}.
Unless otherwise mentioned, we will use `displacements' to imply CRD.

As seen from Fig.~\ref{f1}(a,c,d) for pure, RP-, and CP-systems respectively, $\gamma_{L}(t) \sim t^2$ for small $t_0 \lesssim 1$ indicating the ballistic motion of particles until they feel the presence of each other. Beyond $t_{0}$, Fig.~\ref{f1}(a) depicts that $\gamma_{L}(t)$ in a pure system remains small and independent of $t$ for small $T$ $(\Gamma^{-1}\leq 0.0056)$ -- signalling a solid-like behavior. For large $T$ ($\Gamma^{-1}\geq 0.0157$), $\gamma_{L}(t) \sim t$ shows liquid-like Fickian behavior. For the intermediate $T$, $\gamma_{L}(t)$ shows a changing slope (in log-log scale) beyond $t_{0}$. Interestingly, in the window $0.0056< \Gamma^{-1} < 0.0157$, we find $\gamma_{L}(t) \sim t^{\beta(T)}$ for long times where $\beta(T) < 1$ and this sub-diffusive growth persists until $t\approx 10^{5}$. We plot $\beta$ in aforementioned $T$-window, presented in the inset of Fig. 1(a), which always remains below unity for $10^4\leq t \leq 10^5$.
The $T$-window for the long-time non-Fickian behavior of $\gamma_{L}(t)$ has a large overlap with the solid and partly with the hexatic-phase -- if phase-boundaries are inferred from static correlations~\cite{PhysRevE.109.L062101}.

To develop deeper insight into the unusual motion, we presented in Fig.~\ref{f1}(b) absolute displacements (not cage-relative) of particles in a pure system at low $T$ ($\Gamma^{-1}=0.0112$), where the the system is presumably a solid~\cite{PhysRevE.109.L062101}.
Fig.~\ref{f1}(b) indicates that while most particles jiggle around their equilibrium positions, a (macroscopic) fraction of particles undergo larger displacements $\Delta r(t) \sim a$ and form a long and tortuous string-like path of cooperative motion. The nature of these dynamics is illustrated in Fig.~S2 in SM~\cite{SupplementalMaterial}.
Dynamical heterogeneities of this type are ubiquitous in glassy systems~\cite{PhysRevLett.80.2338, nagamanasa2011confined, hima2015direct, kim2013simulation, zangi2004cooperative, PhysRevLett.80.2338, PhysRevB.60.5721,cui2001dynamical}, in dusty plasma~\cite{chan2009dynamical}, and confined systems~\cite{ash2016spatio}, and are 
attributed to repeated caging and cage-breaking events~\cite{li2020anatomy,pareek2023different}. They are quite unlike the collective oscillations in solids. Interestingly, the non-Fickian $\gamma_{L}(t)$ occurs in the same $T$-window that identifies the cooperative motion temperature regime (CMTR). While cooperative motion in CMTR does occur via caging and cage-breaking even in a pure system, such glass-like events are rife in RP-systems, SM~\cite{SupplementalMaterial}.

How do the dynamics unfold when pinning is introduced? Evolution of $\gamma_{L}(t)$ for RP-systems at different $T$ in Fig.~\ref{f1}(c) indicates a sub-diffusive growth of $\gamma_{L}(t)$ in the entire low-$T$ regime until melting into the liquid.
RP-centers were found to trigger cooperative motion (see Fig. S1 in SM~\cite{SupplementalMaterial}), and CMTR engulfs the entire low-$T$ phase ($\Gamma^{-1} \leq 0.0145$).
The value of $\beta$ in CMTR is typically smaller in the RP-systems,  as shown in the inset of Fig 1(c), and remains largely insensitive within $10^{4} \leq t \leq 10^{5}$. Thus, random impurities strengthen the sub-diffusive dynamics compared to a pure system. All our data asserts $\beta < 1$ within CMTR up to $t\approx 10^{5}$.
Will these non-Fickian dynamics survive for longer times?
While our results point towards such possibilities, more so for RP-systems, our data quality is inadequate for conclusively establishing this. 

Fig.~\ref{f1}(d) shows $\gamma_{L}(t)$ for CP-systems, where dynamical signatures are similar to the pure system. However, the cooperative motion and the eventual Fickian dynamics set in at a higher threshold $T$ because crystallinity is anchored by commensurate pinning.

\textit{Self-part of van Hove correlation}, $G_{s}(r,t)=N^{-1}\langle\sum_{i,j}^{N}\delta \left( r-\vert\Delta r_{i}(t)\vert \right) \rangle$, tracks the spatio-temporal evolution~\cite{PhysRev.95.249,hansen2013theory}, signifies the probability that a particle has traversed a distance $r$ on an average in time $t$. Here $\Delta r_i(t)$ is the CRD of the $i$-th particle. Consistent with the evolution of 
$\gamma_{L}(t)$ in a pure system, the temporal behavior of $2\pi r G_{s}(r,t)$ divides thermal effects into three thermal windows,
shown in Fig.~\ref{f1}(e-g). At a very low $T \sim \Gamma^{-1}=0.0056$, $G_{s}(r,t)$ assumes a narrow Gaussian distribution in (Fig.~\ref{f1}(e)), and remains unaltered in time, signaling solidity. The peak position of $2\pi r G_{s}(r,t)$ marks the width of this Gaussian form.

At an intermediate $T$, a tail develops in $2\pi r G_{s}(r,t)$ with increasing weight as time elapses (Fig.~\ref{f1}(f)), reducing its peak height. The peak position, however, remains unchanged. The emergence of this long tail and the unaltered peak position fully overlap with the CMTR. At higher $T$ (Fig.~\ref{f1}(g)), $G_{s}(r,t)$ broadens with $t$ causing a drifting peak of $2\pi r G_{s}(r,t)$ to larger $r$ with falling intensity -- signaling liquidity.

Turning to RP-systems, we uncover that CMTR swamps the entire low-$T$ regime until liquidity sets in for $\Gamma^{-1} \leq 0.0145$, and representative $G_{s}(r,t)$s at low-$T$ are shown in Fig.~\ref{f1}(h). The liquid-like behavior at high-$T$ in RP- and CP-systems (not shown separately) is qualitatively similar to the pure system, modulo the different melting $T$. The low-$T$ trend of $2\pi r G_{s}(r,t)$ in a CP-system is identical to that in a pure system. However, solidity survives up to a higher $T$. Its behavior in CMTR (Fig. ~\ref{f1}(i)) features an expected long tail along with an unaltered peak position.

The cooperative motion in CMTR is analyzed by the distinct part of the van Hove correlation, denoted as $G_{d}(r,t)$. It measures the propensity of finding a different particle in time $t$ at a distance $r$ from where a particle was initially located. Fig. ~\ref{f1}(j) shows $G_{d}(r,t)$ in absolute coordinate (not cage-relative) for a pure system at $\Gamma^{-1}=0.0112$ (solid phase). The emerging peak in $G_{d}(r\rightarrow 0)$ rises with $t$ signifying particles occupying the location from where another particle moved out, setting up a cooperative motion. Note that the sharpness of the peaks of $G_{d}(r,t)$ for $r \ge a$ alters only marginally, even at the largest times, confirming solidity. The emerging peak in $G_{d}(r\rightarrow 0)$ was found only within CMTR both in the presence and absence of pinning.~\footnote{For a probabilistic interpretation of $G_{d}(r,t)$, i.e. conservation of area under its trace, it should be multiplied by  $2\pi r$, similar to the plots for $G_{s}(r,t)$. However, our focus is on the emergence of the peak in $G_{d}(r\rightarrow 0)$, and we disregarded the factor $2\pi r$ for visual clarity.} We now quantify the departure from the Gaussian form of displacements in CMTR.

\textit{Non-Gaussian parameter (NGP)}, is defined as $\alpha_{2}(t)=({\langle \Delta r^{4}(t)\rangle}/{2\langle \Delta r^{2} (t)\rangle^{2}})-1$. It measures the extent to which a distribution of CRD, $\Delta r(t)$, deviates from a Gaussian form with time. Fig.~\ref{f3}(a) elucidates the time evolution of NGP for a pure system.
For low-T ($\Gamma^{-1}\leq 0.0056$), $\alpha_{2}(t)\approx 0$, as expected for pure solids, where particle displacements in phonon modes are Gaussian distributed. The cooperative motion sets in for $0.0056< \Gamma^{-1}<0.0157$ and NGP shoots up sharply past $t/t_0 \approx 10^2$ reaching a maximum at $t_{\rm max}(T)$ and decays thereafter. Our results capture a broadly decreasing trend of $t_{\rm max}$ and $\alpha_{2}(t_{\rm max})$ with $T$~\cite{zangi2004cooperative}.
The NGP for a pure system is shown in CMTR as an inset of Fig.~\ref{f3}(a) for $t \leq 10^5$. While $\alpha_{2}(t)$ falls monotonically beyond $t_{\rm max}$, reaching down to a sixth of its peak value, $\gamma_{L}(t)$ remains sub-diffusive, and cooperative motion persists in CMTR at these long times, in contrast with earlier findings~\cite{zangi2004cooperative, kim2013simulation,van2015dynamical}.
For larger $T$ ($\Gamma^{-1}\geq 0.0157$), the pure system transits to the liquid phase, depleting $t_{\rm max}$ and $\alpha_{2}(t_{\rm max})$ rapidly.

Turning to $\alpha_{2}(t)$ for RP-systems in Fig.~\ref{f3}(b), we find $\alpha_{2}(t)\neq 0$
down to the lowest $T$, denying the existence of an ideal solid. $\alpha_{2}(t)$ in CMTR for RP-systems shown as the inset of Fig.~\ref{f3}(b) highlights that $\alpha_{2}(t>t_{\rm max})$ decreases steadily. Yet, $\beta(T) < 1$ in CMTR shows little variation for $10^4 \leq t \leq 10^5$. Thus, our results are in contrast with sub-diffusive $\gamma_{L}(t)$ for $t < t_{\rm max}$ followed by Fickian growth at longer times predicted in the context of different colloidal systems~\cite{ zangi2004cooperative, kim2013simulation,PhysRevLett.118.158001, kim2020dynamical, wang2010two, van2015dynamical}, dipolar systems~\cite{PhysRevE.102.033205} and other heterogeneous media~\cite{horbach2017anomalous}. 
NGP in the inset of Fig.~\ref{f3}(b) shows that both $t_{\rm max}$, and $\alpha_{2}(t_{\rm max})$ decreases with $T$ in RP-systems while they showed little variation for a pure system. However, the CMTR is far narrower in pure systems for justified comparisons. Beyond melting into a liquid for $\Gamma^{-1} > 0.0145$, NGP features a standard liquid-like behavior -- the liquidity does not qualitatively distinguish the clean and pinned systems. The behavior of NGP in CP-systems (Fig.~\ref{f3}(c)) is qualitatively similar to those in a pure system, albeit with modified boundaries of $T$-windows.

\textit{Measure of the non-Gaussian motion:}
The traces of $G_s(r,t)$ in CMTR features a largely monotonic decay of the form $\sim e^{-lr^k}$~\cite{ash2016spatio} for $t>t_0$ (Fig.~S4 in the SM~\cite{SupplementalMaterial}).
Here, $k=2$ signals a diffusive motion, while $k=1$ implies an exponential fall of $G_s(r,t)$ with $r$. A generic glassy dynamics yields $k=1$~\cite{PhysRevLett.99.060604}.
We extracted the time dependence of $k$ for several $T$ following the prescription of Ref.~\onlinecite{ash2016spatio} (See  SM~\cite{SupplementalMaterial}). Our results are presented for pure, RP- and CP-systems in Fig.~\ref{f3}(d, e, f).

We found $k(t\lesssim 5t_0)\approx 2$ for all $T$ at initial times. We find $k(t)\approx 2, \forall t$ at low-$T$ ($\Gamma^{-1}\leq 0.0056$) in pure systems (Fig.~\ref{f3}(d)). In CMTR ($0.0056 <\Gamma^{-1}<0.0157$), $k(t)$ decreases rapidly reaching $k(t)<1$ as time increases. However, they show a final upturn near the largest times of our study. This is consistent with the results of NGP in CMTR. The high-$T$ results in the pure system feature a non-monotonic behavior of $k(t)$ in which $k$ initially decreases from $2$ to nearly $1$ before it turns around and steadily rises again towards $k=2$ at longest times.

Fig.~\ref{f3}(d) shows that $k(t)$ depletes from $2$ to a nearly $t$-independent value ($\approx 1.5$) past $t/t_0 \approx 10^2$ for RP-systems, even for the lowest $T$.
Interestingly, this finding is consistent with the absence of true solidity in RP-systems. The hexatic correlations remained strong at low $T$~\cite{PhysRevE.109.L062101}.
Together with slow structural relaxation consisting of caging and cage-breaking events akin to glasses, the strong hexatic correlations~\cite{PhysRevE.109.L062101} in the low-$T$ phase of the RP-system makes it tempting to identify this phase as a `hexatic glass' -- an exotic phase proposed in the context of vortex phases of type-II superconductor~\cite{PhysRevB.40.11355, chudnovsky1991orientational, toner1991orientational}. For $T$ $\sim \Gamma^{-1}\leq 0.0145$, $k(t)$ keeps decreasing monotonically, with $t$ reaching as low as $k\approx 0.5$ with little sign of rise at longer times. A further increase of $T$ restores the usual liquid-like behavior.

A similar analysis for the CP-systems in Fig.~\ref{f3}(f) yields similar conclusions as for pure systems, though the boundaries between the low-$T$ ($\Gamma^{-1}\leq 0.0140$) solid-like, intermediate-$T$ CMTR ($0.0140<\Gamma^{-1}\leq 0.0182$), and the high-$T$ liquid-like behaviors are different. Still, these differences are consistent with the findings from NGP.

\textit{Conclusions:}
Analyzing dynamics across 2D melting of Gaussian-Core particles in pure and disordered environments, we found unusual motional footprints at low temperatures, termed CMTR. 
\begin{figure}[h]
\centering
\includegraphics[width= 9.5 cm,height=3.8 cm]{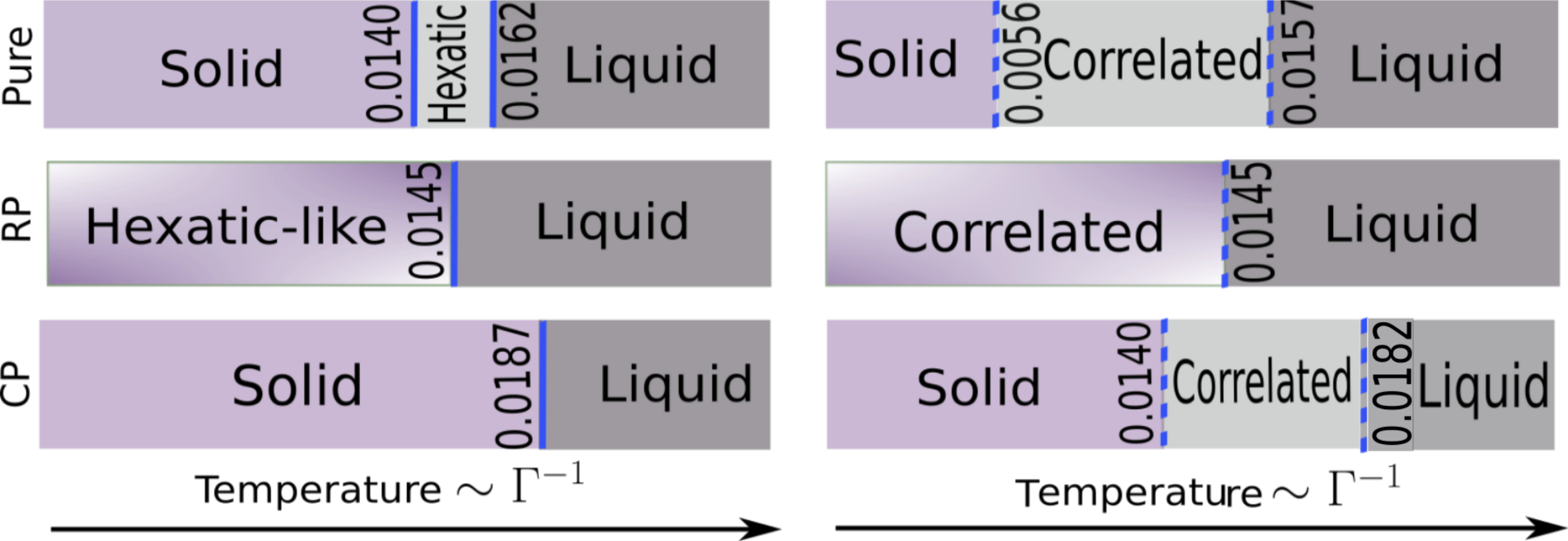}
\caption{{\bf A schematic representation of phases} associated with 2D-melting in pure and impure systems is shown from the behavior of static and dynamic correlations. The thermodynamic phases and boundaries are shown on the left panels. The values of $T \sim \Gamma^{-1}$ are marked on the phase boundaries~\cite{PhysRevE.109.L062101}, indicating that the 2-step BKTHNY melting in the pure system is obscured by impurities.
The right panels exhibit the dynamical behaviors across the melting, illustrating that the motional footprints differ from the thermodynamic phases and their boundaries. We use dotted lines to demarcate $T$-windows of the dynamical patterns.}
\label{f4}
\end{figure}
These include sub-diffusive $\gamma_{L}(t)$, non-Gaussian displacements, and dynamic heterogeneity arising from the cooperative motion of certain particles along tortuous string-like paths.
Our usage of CRD cuts off the long wavelength collective motion, which otherwise causes $\gamma_L$, expressed in standard coordinates, to diverge even in a solid at finite $T$; the structural relaxation, typical for glasses, still occurs.
Although familiar to glasses, such cooperative motions, even in pure systems, are unanticipated in equilibrium phases under consideration. The effect of such atypical motion is amplified in the presence of uncorrelated disorders, where they are found down to the lowest temperatures, denying solidity. In contrast, impurities at random vertices of underlying triangular lattice anchors solidity.
Our results of dynamical behavior in CRD were compared with observables using standard displacements. While there are qualitative similarities, finer distinctions between them
are found consistent with those reported in  Ref.~\onlinecite{illing2017mermin}. While these temporal behaviors should not be confused with dynamical phase transitions, their comparison with different phases of dynamical 2D systems (e.g., active matter~\cite{klamser2018thermodynamic}) would be interesting. Intriguingly, the thermodynamic phases and their boundaries (Fig.~\ref{f4}) differ from the boundaries between distinct temporal behaviors.

\textit{Acknowledgement: \textemdash }
The computations were performed using Kepler, Dirac and CQM clusters at IISER Kolkata. SD gratefully acknowledges the support from Prime Minister's Research Fellowship (PMRF), India. AG acknowledges SERB research grant No. MTR/2022/000946 from Govt of India.

\vspace{1.0 cm}
\bibliography{ref}
\newpage
\pagebreak
\clearpage  
\vfill
\onecolumngrid  
\begin{center}
\textbf{\LARGE Supplementary material for ``Cooperative motion in equilibrium phases across two-dimension melting in pure and disordered systems''}
\end{center}

\vspace{0.7 cm}

In order to support the key conclusions reported in the main manuscript, we have included additional results in the supplementary materials (SM) below. 
\setcounter{figure}{0}
\renewcommand{\thefigure}{S\arabic{figure}}
\section{Parameters and details of simulation}
We used molecular dynamics simulation~\cite{frenkel2023understanding} within the canonical ensemble to study these systems.
The dimensionless interaction parameter for our model is defined by $\Gamma = \exp(-A/\rho)/K_{B}T$, with 2D particle density $\rho$ and the thermal energy $K_{B}T$. Here, A is a constant of the order of unity. The geometry of the spatial distribution of the particles determines its value.
Here, we take $A=\sqrt{3}/2$, the value at $T=0$ where the constituent particles arrange themselves in a triangular lattice.
The simulations were carried out using LAMMPS~\cite{PLIMPTON19951} with periodic boundary conditions and a simulation box size of $L_{x}= \frac{2}{\sqrt{3}} L_{y}$. The $L_{x}$ value is modified to maintain a constant density, $\rho$ of particles for all cases reported in the main text ($\rho = 0.628$). Maintaining a fixed density allowed us to tune $\Gamma$ by adjusting $T$ alone. We conducted $2 \times 10^7$ MD steps for our simulations with a time step of $\delta t = 0.005$. Subsequently, we use $2 \times 10^6$ MD steps ($t = 10000$) with a sampling time window of $t=0.1$, recorded after the equilibration run. We used dimensionless parameters: $t' = t \sqrt{1/m \sigma^2}$ and $E' = E$, where $m$ represents the mass of each particle.  The desired temperatures are maintained via the Berendsen thermostat (velocity rescaling)~\cite{frenkel2023understanding}.
We ensured accurate equilibration before collecting statistics by studying the distribution of velocities of the particles (expectedly a Maxwell-Boltzmann distribution in thermal equilibrium) as well as the independence of the temporal correlation of the observables on time origins, i.e. $\zeta(t_{1},t_{2}) = \zeta_(t_{1}-t_{2})$, where $\zeta$ is any temporal correlation function defined between two time points $t_{1}$ and $t_{2}$. In our notation, $\langle \cdots \rangle$ denotes an average over the number of particles and MD configurations and over pinning realizations for its given concentration.
\section{Trajectories and Displacements}
We present in Fig.~S1 the trajectories and displacements of the particles in pure and pinned systems. Fig.~S1(a) shows the trajectories of particles for a specific realization of dynamics up to $t \le 10^{4}$ at low $T$ ($\Gamma^{-1} = 0.0112$) which was identified as a solid by static correlations~\cite{PhysRevE.109.L062101}.  Fig.~S1(b) represents the particle displacements corresponding to Fig.~S1(a) in bare (or regular) coordinates (not cage-relative). Here, the initial positions of the particles are denoted by red dots. At the same time, the lengths of the connecting blue lines indicate the magnitudes of the $i$-th particles’ displacements at a specific time $t$, $\Delta r_{i}(t)=r_{i}(t)-r_{i}(0)$. It is seen that the majority of particles only rattle around their equilibrium positions. However, a (macroscopic) subset of particles demonstrates a long-range motion in a long and tortuous path of spatially correlated motion. A low-$T$ regime rife with such a motional signature is termed a CMTR from temporal characterization. We also define the cage-relative displacements of the $i$-th particle, $\Delta r_{i,{\rm rel}}(t)=\Delta r_{i}(t) - \Delta r_{i+\delta}(t)$, i.e., the cage-relative displacements (CRD) of particles with respect to their first neighbors at $\delta$. All our results in the main text, except for $G_{d}(r,t)$, are presented using cage-relative coordinates.
The {\it standard} displacements, corresponding to Fig.~S1(a,b), are shown as Fig.~1(b) in the main manuscript. The cage-relative displacements $\Delta r_{i,{\rm rel}}(t)$ displays crystal-like dynamics more effectively and suppresses the collective bodily motion. Also, notice that the excursions of the particles are notably reduced in $\Delta r_{i,{\rm rel}}(t)$. Nevertheless, certain particles, typically related among themselves by one of the nearest neighboring relationships, still demonstrate long-range spatially correlated motion. Fig.~S1(c,d) represents particle trajectories and cage-relative displacements, similar to Fig.~S1(a,b), but for a specific realization of RP-impurities. The spatially correlated motion of the particles in a long and tortuous path abounds in RP-systems, compared to the pure system at the same temperature. Such motional footprints are found in RP-systems down to the lowest $T$ of our study. This is illustrated in Fig.~S1(e) by showing trajectories at $\Gamma^{-1}=0.0056$.
Fig.~S1(f) represents similar trajectories for a CP-system at $\Gamma^{-1}=0.0140$. In this case, the near-perfect `solidity' demonstrates that the commensurate pinning anchors the solidity up to a much larger $T$.

\begin{figure}[h]
\centering
\includegraphics[width= 15.0 cm,height=22.5 cm]{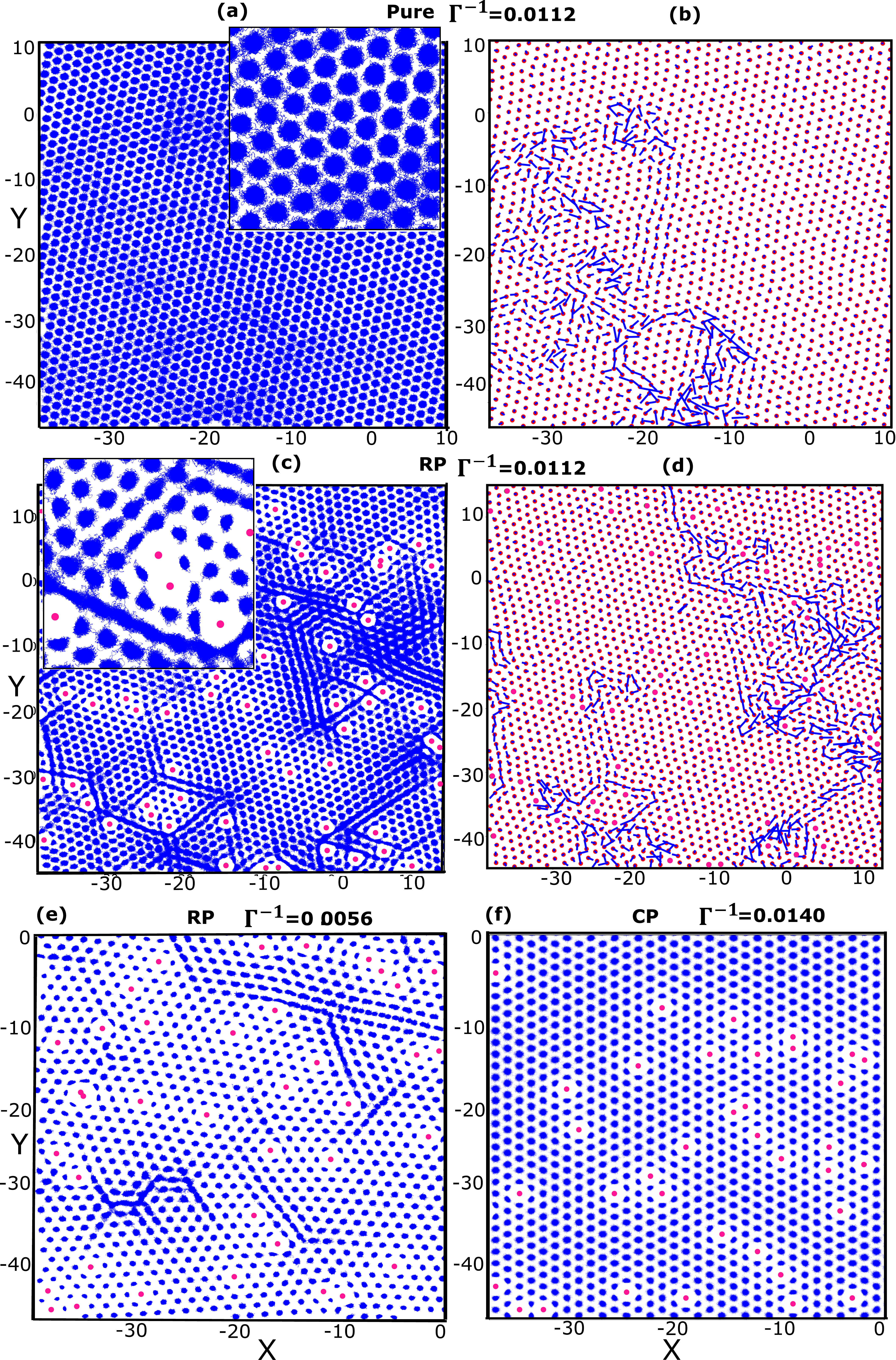}
\caption{A section of the simulation box showing the trajectories and displacements of the particles for visual clarity. Panel (a) and (b) show the trajectories and cage-relative displacements of the particles in the pure system for a specific realization of dynamics up to $t \le 10^{4}$ at low $T$ ($\Gamma^{-1} = 0.0112$). In cage-relative displacements, the red dots indicate the initial positions of the particles, and the connecting blue lines represent the displacements of the particles from their initial position. On the other hand, panels (c) and (d) show the same for the RP system. Here, the center of pinned particles is represented by the pink dots. The insets in panels (a) and (c) present magnified views of the trajectories within the bounding box $x \in [-25.0,-15.0]$ and $y\in [-42.5,-32.5]$ for the clean system and $x\in [-30.0,-20.0]$ and $y\in[-40.0,-30.0]$ for the RP system. This enhances the visual clarity of the complex trajectories of individual particles. Panel (e), and (f) show the trajectories of the particles for the RP and CP system at $\Gamma^{-1}=0.0056$ and $\Gamma^{-1}=0.0140$ respectively.}
\label{s1}
\end{figure}

\section{Analysis of the spatially correlated motion of a macroscopic fraction of particles at low temperatures}

Here, we demonstrate the nature of the spatially correlated motion of a macroscopic fraction of particles in a string-like path at a low $T$ solid phase in a pure system. Focusing on a typical realization of displacements in a section of Fig.~1(b), as indicated by the coordinates of its bounding box, we reproduce the displacements in this region in Fig.~S2(a) where we mark six particles by the number $1$ to $6$ and tag their initial position by different colored dots. In contrast, the initial positions for the remaining particles remain red. The six particles are so chosen that they participate in two separate spatially correlated motion paths (three each -- particles $1$, $2$, and $3$ in one, while $4$, $5$, and $6$ in the other). The trajectories of these six particles are enlarged in Fig.~S2(b-g), respectively, along with one solid line in each representing the net displacement. We observe that these particles jiggle around their equilibrium position for a certain time and then {\it jump} by a distance $\sim a$ (average lattice spacing at the lowest $T$) and spend time there before jumping either to the previous (reversible motion) or to a new location (irreversible motion). Such occasional jumps can be captured by looking into the instantaneous position, $r_i(t)$, as shown in Fig.~S2(h), where $i$ represents the particle index. We see that within a sufficiently small time window (hence, we call them ‘jumps’), these particles change their positions, moving by a distance of $a$. This motion can be regarded as the cage breaking process. These jumps are not always exact, as we see that the jumps of the particles $\#1$, $\#2$, and $\#3$ in the first of the two motion groups are sharper than, say, the particle $\#4$ in the second motion group. This is also reflected in the trajectory of particle $\#4$, which, upon leaving its initial equilibrium position, samples a few others nearby before settling close to its final equilibrium position and continues to jiggle around it. Nevertheless, the time spent jiggling around an equilibrium position is significantly larger than the time of the jump, as illustrated by Fig.~S2(h). Similar motional footprints have also been found for the pinned systems in the corresponding CMTR.

\begin{figure}[h]
\centering
\includegraphics[width= 15.07 cm,height=24.5 cm]{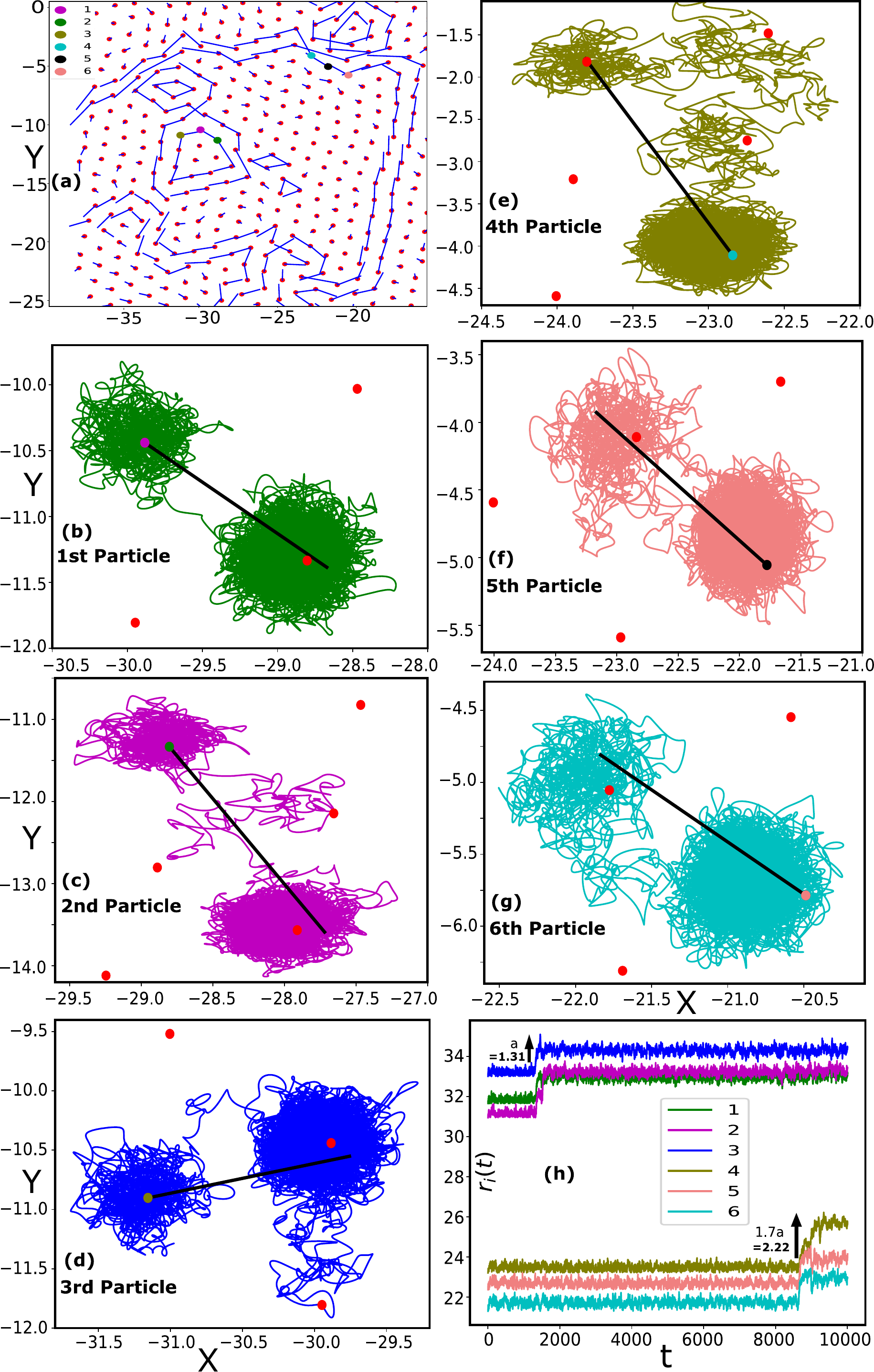}
\caption{Nature of spatially correlated motion of macroscopic fraction of particles is shown in panel (a). Among them, we marked the initial positions of six particles with different colors whose motions were analyzed. Panels (b-g) show the individual trajectories of those six particles. The $t$-dependence of the displacements of particles from their equilibrium position is shown in Panel (h).}
\label{s2}
\end{figure}

\section{Contrasting time evolution of $G_s(r,t)$ at low $T$ in RP- and CP-systems}
In the main text, we claimed that the dynamical behavior of particles grants a characteristic evolution of $G_{s}(r,t)$, which carries the signature of the CMTR in the entire low-$T$ region until the RP-systems turns into a liquid. While this was demonstrated in the main manuscript for $\Gamma^{-1}=0.0112$, we show in Fig.~S3(a) that such signal, though weaker, is present for the lowest $T$ of our study ($\Gamma^{-1}=0.0056$). The main panel (a) illustrates the slow and gradual fall of the peak height of $2\pi r G_{s}(r,t)$ with $t$ keeping the peak position unaltered, while the inset emphasizes the gradual lengthening of the tail with $t$. Interestingly, a faint secondary peak can also be discerned in the extended tail at a large time, similar to what was observed in Fig.~2(h) in the main manuscript.

In addition, we also show the behavior of $G_{s}(r,t)$ for the CP-system in Fig.~S3(b) at a temperature ($\Gamma^{-1}=0.0140$) $2.5$ times higher than that of the panel (a). It exhibits a Gaussian-like behavior of $G_{s}(r,t)$ for all $t$. This suggests the system retains a solid-like behavior up to high temperatures, attributed to the pinned particles commensurate with the underlying perfect lattice.

\begin{figure}[h]
\centering
\includegraphics[width= 16.0 cm,height=5.8 cm]{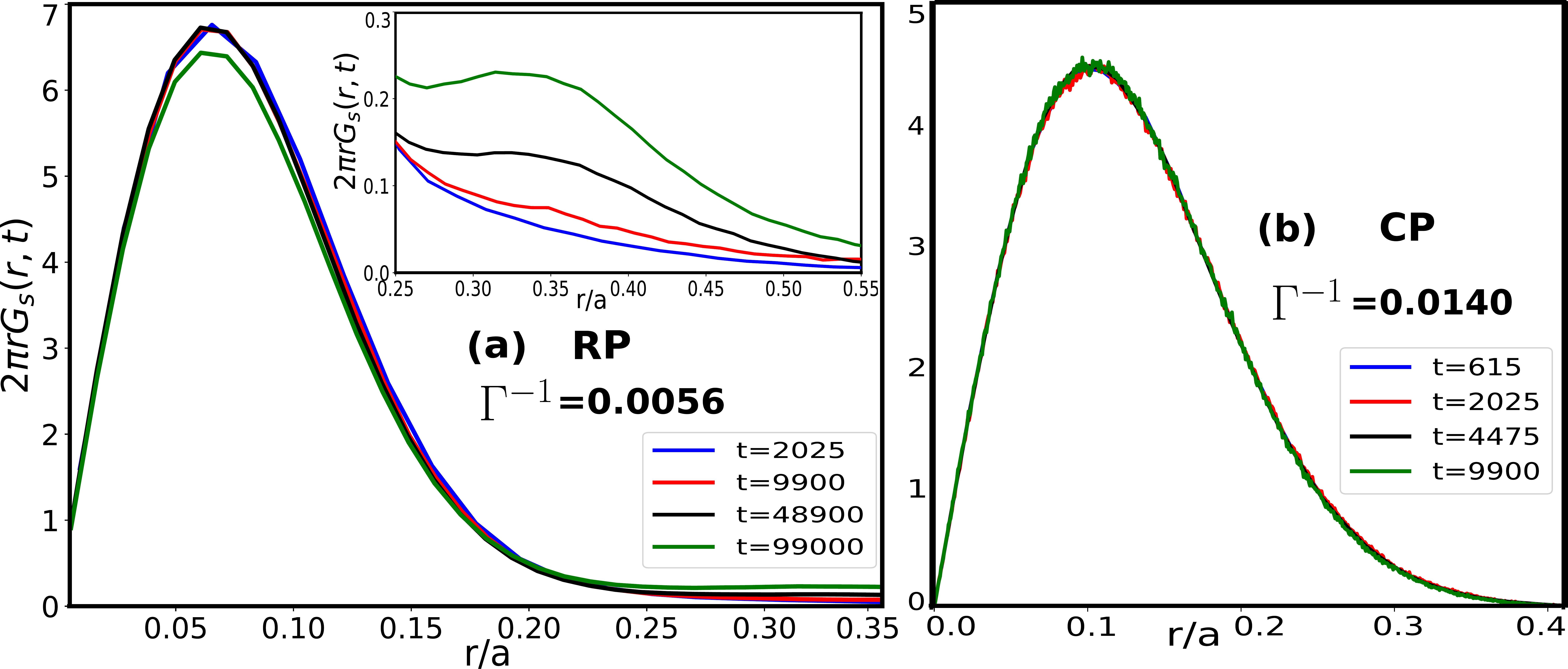}
\caption{The self-part of the van Hove correlation function for the RP and CP systems is displayed in panel (a) and (b) at $\Gamma^{-1}=0.0056$ and $\Gamma^{-1}=0.0140$ respectively. The inset in panel (a) shows the long-tail behavior of $G_{s}(r,t)$ at large $r$, indicating the presence of cooperative motion even at the lowest $T$ of our simulation. In contrast, the CP system anchors the solidity even at the higher temperature. }
\label{s3}
\end{figure}

\section{Extraction of the exponent $k(t)$}
To quantify the degree of non-gaussianity, we analyze the large-$r$ decay of $G_{s}(r,t)$ by fitting it to a form $\sim {\rm exp}(-lr^{k})$, where $k\equiv k(t)$ is a time-dependent quantity. The small-$r$ behavior of $G_{s}(r,t)$ was always found Gaussian for all $t$ and $T$. Thus, we employed two functional dependencies to fit the molecular dynamics (MD) data: $G_{s}^{\rm small}(r,t) \sim e^{-r^{2}/c}$ for small value of $r$ and $G_{s}^{\rm large}(r,t) \sim e^{-lr^{k}}$ for large value of $r$ for a wide range of $\{t, T\}$. Here, we display the exponent $k(t)$ with $t$ for RP-system at  $\Gamma^{-1}=0.0112$ in Fig.~S4. It is evident that $k(t)$ exhibits a continuous and consistent decrease as $t$ increases, and the value of $k$ approaches $\approx 0.5$ at the highest $t$ of our simulation. Note that for the temperature under consideration (falling in the window of CMTR), we find no indication of a long-time increase of $k(t)$! The insets illustrate the fitting procedures to extract the exponent $k(t)$ at two distinct time points -- $t=7.9$ and $t=9900.0$, respectively. In this representation, the black dots correspond to the $G_{s}(r,t)$ data, while the green and red lines represent the fitting functions $G_{s}^{\rm small}(r,t) \sim e^{-r^{2}/c}$ and $G_{s}^{\rm large}(r,t) \sim e^{-lr^{k}}$ respectively, all expressed in a log-scale.
\begin{figure}[h]
\centering
\includegraphics[width= 9.5 cm,height=6.5 cm]{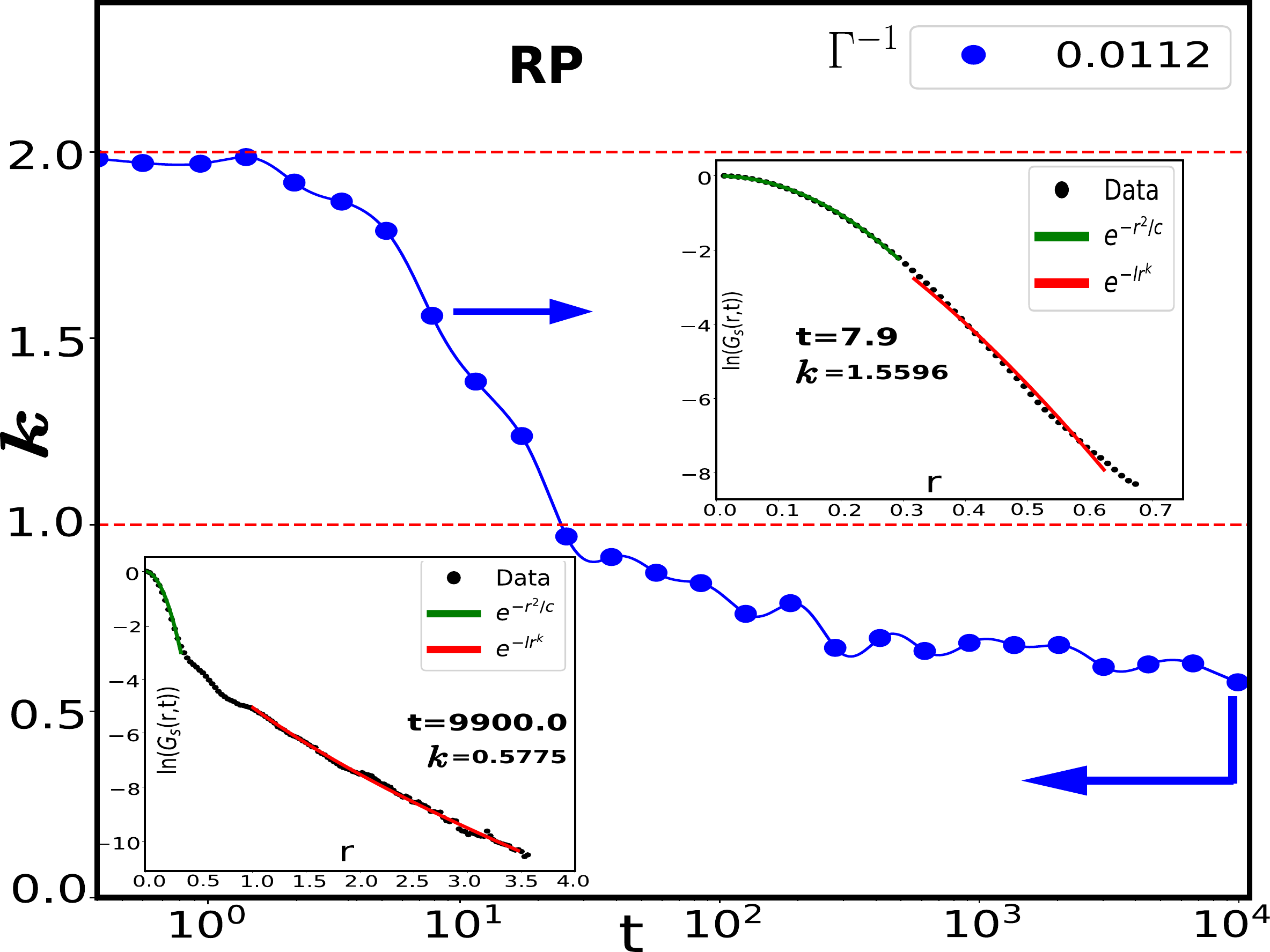}
\caption{The behavior of $k(t)$ as extracted for the RP-system at a temperature $\Gamma^{-1}=0.0112$ from a fitting procedure explained in the text. The system exhibits cooperative motion of particles in string-like paths at this value of $\Gamma^{-1}$.}
\label{s4}
\end{figure}

\section{Comparison between regular and cage-relative co-ordinate}
In this section, we highlight the differences between dynamics recorded in regular coordinates and the cage-relative coordinates. In particular, we emphasize how such differences reflect on the physical observables, such as $G_{s}(r,t)$, more specifically in the CMTR.
In regular coordinates, $G_{s}(r,t)$ represents the probability that a particle traversed a distance $r$ on the average in time $t$,
whereas, in cage-relative coordinates, it represents the probability that a particle traversed a distance $r$ to its first neighbors in time $t$.

Our results of $G_{s}(r,t)$ in both these coordinates are shown in a semi-logarithmic plot in Fig.~S5. Here, panels (a,b), (c,d), and (e,f) show the $r$ dependence of $G_{s}(r,t)$ for regular and cage-relative coordinates, respectively for pure, RP- and CP-systems at $\Gamma^{-1}=0.0146, 0.0134$ and $0.0182$ respectively, for a few different $t$'s.

While the differences are minimal between the two at short times -- both showing a Gaussian trend as expected, as time progresses, additional peaks emerge for $r/a \approx 1, 2, 3$, etc., when $G_{s}(r,t)$ is expressed in regular coordinates. In addition to featuring these peaks, $G_{s}(r,t)$ also develops extended tails for long times.
However, in cage-relative coordinates, these peaks are essentially invisible, though some peak-like features with reduced intensity can be discerned for the pure system. However, the long tail of $G_{s}(r,t)$ remains prominent even in the cage-relative coordinates. Such a disparity in the behavior of $G_{s}(r,t)$ may occur due to the inhibition of collective modes in the cage-relative coordinate.

\begin{figure}[h]
\centering
\includegraphics[width= 12.0 cm,height=18.0 cm]{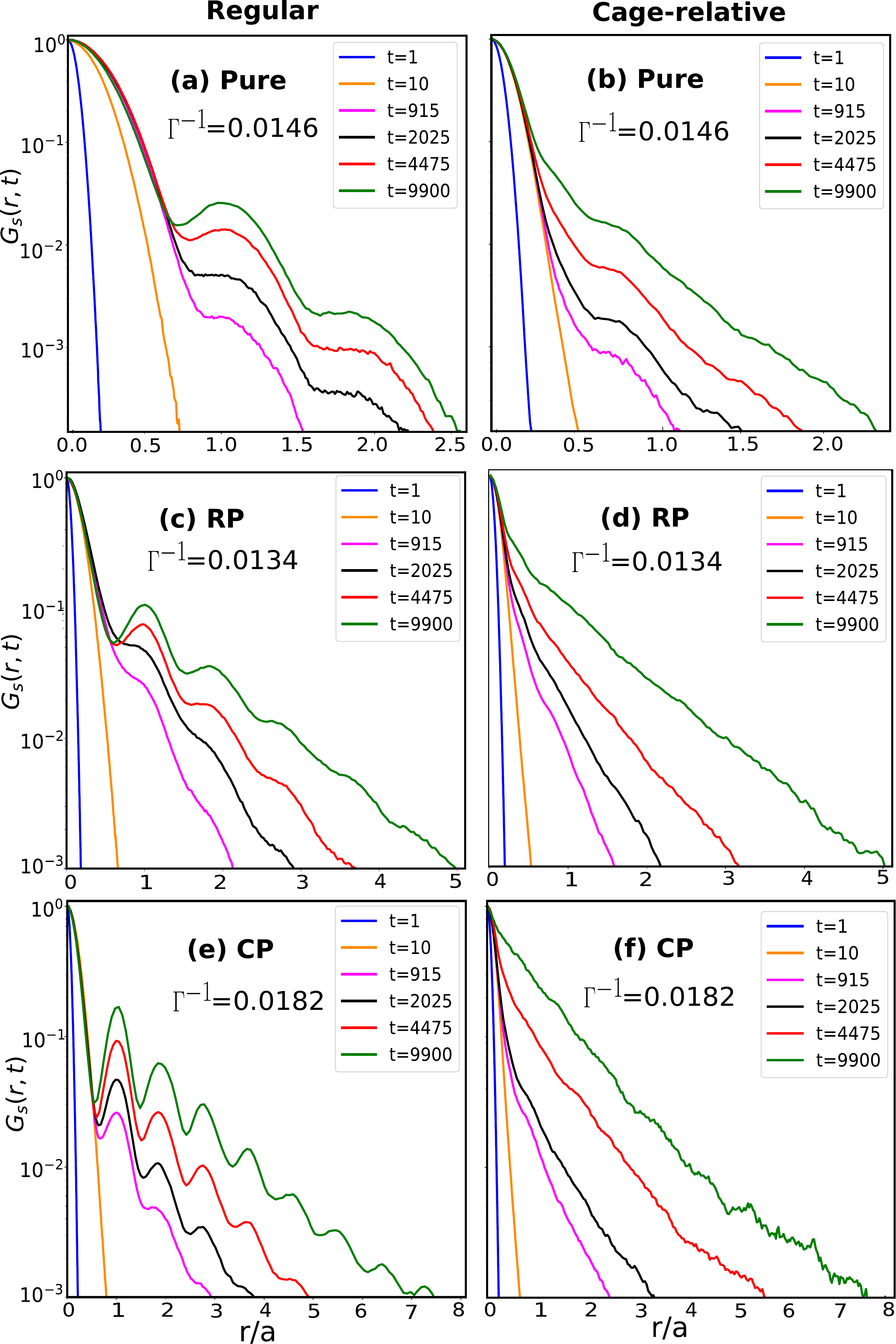}
\label{s6}
\caption{The comparison of $G_{s}(r,t)$ in the CMTR, expressed in the regular coordinates (left panels) and cage-relative coordinates (right panels) for pure (panel (a) and (b)), RP- (panel (c) and (d)), and CP-systems (panel (e) and (f)). The occurrence of its multi-peak profile for long times, along with the long tail in regular coordinates, is qualitatively distinct from when it is plotted in the cage-relative coordinate, which features only the long tail of the distribution.}

\end{figure}
\end{document}